\documentclass[fleqn,10pt]{wlscirep}
\usepackage[utf8]{inputenc}
\usepackage[T1]{fontenc}

\usepackage{dcolumn}
\usepackage{bm}
\usepackage{graphicx}
\usepackage{epstopdf}
\usepackage{wrapfig}
\usepackage{array} 
\usepackage{listings}
\usepackage[para,online,flushleft]{threeparttablex}
\usepackage{booktabs,dcolumn}

\usepackage{textpos}
\usepackage{booktabs}
\usepackage{multirow,bigdelim}
\usepackage{float}

\usepackage{xcolor}
\definecolor{pastelgray}{rgb}{0.81, 0.81, 0.77}
\definecolor{beaublue}{rgb}{0.9, 0.9, 0.93}
\usepackage{orcidlink}

\usepackage{xr-hyper}
\usepackage{hyperref}
\usepackage{multibib}
\newcites{methods}{References for Methods}

\usepackage{bm} 

\newcommand{\alphab}{\bm{\alpha}}
\newcommand{\betab}{\bm{\beta}}

\newcommand{\deltab}{\bm{\delta}} 

\newcommand{\Db}{\bm{D}}

\newcommand{\yb}{\bm{y}}    

\newcommand{\omegab}{\bm{\omega}} 

\title{Local Bayesian Dirichlet mixing of imperfect models}

\author[1,$\dagger$]{Vojtech Kejzlar \orcidlink{000-0002-1001-011X}}
\author[2]{L\'eo Neufcourt \orcidlink{0000-0002-0502-4429}}
\author[3,*]{Witold Nazarewicz \orcidlink{0000-0002-8084-7425}}
\affil[1]{Mathematics and Statistics Department, Skidmore College,
Saratoga Springs,
New York 12866, USA}
\affil[2]{FRIB Laboratory, Michigan State University, East Lansing, Michigan 48824, USA}
\affil[3]{Department of Physics and Astronomy and FRIB Laboratory, Michigan State University, East Lansing, Michigan 48824, USA}

\affil[$\dagger$]{vkejzlar@skidmore.edu}
\affil[*]{witek@frib.msu.edu}

\begin{abstract}
To improve the predictability of complex computational models in the experimentally-unknown domains, we propose a Bayesian statistical machine learning  framework utilizing the Dirichlet distribution that combines results of several imperfect  models. 
 This framework can be viewed as  an extension of
Bayesian stacking. To illustrate the method, we study the ability
of Bayesian model averaging and mixing techniques to mine nuclear masses.
 We show that  
 the global and local mixtures of models
 reach excellent performance on both
prediction accuracy and uncertainty quantification and are preferable to classical Bayesian model averaging. Additionally, our statistical analysis indicates that improving model predictions through mixing rather than mixing of corrected models leads to more robust extrapolations.
\end{abstract}
\begin{document}

\flushbottom
\maketitle
\thispagestyle{empty}

\section*{Introduction}

When considering predictions or extrapolations of physical quantities into unknown domains, a reliance on a single imperfect theoretical model can be misleading. To improve the quality of extrapolated prediction, it is in fact advisable to use several different models and mix their results \cite{Hoeting1999,Was00,Fragoso2018,BANDmanifesto}.
In this way, the ``collective
wisdom'' of several  models could be maximized
by providing the best prediction rooted in the most current
experimental information. To carry out the mixing,
statistical machine learning (ML) methods, with their ability to capture the local features, are tools of choice. Specifically, Bayesian Model Averaging (BMA) can be used to combine multiple models to produce more reliable  predictions since it is the 
natural Bayesian framework to account for the uncertainty
on the model itself \cite{Hoeting1999,Was00,Fragoso2018}.
In absence of another established methodology, the application of BMA to scenarios
where several models compete to describe 
the same phenomenon
has been relatively widespread in such diverse fields
 as weather forecasting
\cite{MWR2011}, political science \cite{Montgomery2010}, transportation \cite{Zou2012}, and nuclear physics \cite{Neufcourt2019,Everett2021}.

It is important to remember, however, that BMA relies on the assumption
that one of the proposed models is the true model
(i.e., a model that perfectly describes the physical reality),
which is clearly inappropriate when dealing with complex systems and approximate modeling. 
In practice, it often happens that
none of the competing state-of-the-art models 
can be dominated by the others, 
in the sense that each model does something better than the others.
In such a setup, models should not be viewed as exclusive but as complementary,
and BMA seems theoretically ill-grounded.
In addition, in the standard implementations of the BMA, the weights are global, i.e., they are constant over the input domain,
and thus unable to catch local model preferences. 

Besides BMA, there exist other methods to combine results of different models.
In fact, combining models has been the subject of much research in ML that has led to the development of the topical ``ensemble learning methods'' 
(bagging and boosting). 
These methods remain in their spirit and purpose very close to BMA and typically do not fix the inadequacies mentioned above: 
their goal is to identify the best performing model given a set of models.
See Ref.\cite{Clemen1989} for review of additional approaches. 

In this work, we develop and apply the Local Bayesian Model Mixing (LBMM), an extension of Bayesian stacking \cite{LeClarke2017, StackingGelman18, StackingGelman22, Semposki22, yannotty2023model}, for managing competing models. Under the Bayesian stacking framework, one assumes that the true model is a linear combination of  the models instead of being one of the models. The extrapolations are thus obtained via a direct mixture of the models, as compared to the mixture of posterior distributions under the standard BMA. 
Unlike the BMA weights which reflect the fit of a statistical model to data, 
\emph{independently of the set of available models} except for normalization, 
the weights based on model mixing or stacking reflect the model's contribution to the final predictions \cite{StackingGelman22}. 
The LBMM used in this study makes the use of Dirichlet distribution to infer stacking weights which hierarchically depend on the model input space and thus highlight the local fidelity of theoretical models. Additionally, the LBMM framework well captures uncertainties of individual models and their mixing weights through the proposed hierarchical structure. Below, we first present the general LBMM framework followed by a pedagogical case of global mixture of models that corresponds to classical Bayesian stacking. Subsequently, we let the model weights vary across the model input space and consider several hierarchical Bayesian models based on the Dirichlet distribution. 

As an example, we apply the new method to predicting nuclear mass, or binding energy, which is the basic property of the atomic nucleus. Since we consider BMA to be a point of comparison for our LBMM methodology, we briefly review the general predictive framework of BMA in Methods section. The binding energy determines nuclear stability as well as nuclear reactions and decays. Quantifying the nuclear binding is important for many nuclear structure and reaction questions, and for understanding the origin of the elements in the universe. The astrophysical processes responsible for the nucleosynthesis in stars often take place far from the valley of beta stability, where experimental masses are not known. In such cases, missing nuclear information must be provided by extrapolations. Accurate values for nuclear  masses and their uncertainties beyond the range of available experimental data are also used in other scientific fields, such as atomic and high-energy physics, as well as in many practical applications. In order to improve the quality of model-based predictions of masses of rare isotopes far from stability, ML approaches can be used that utilize experimental and theoretical information. A broad range of ML tools have been used to mine unknown nuclear masses, including Gaussian processes (GPs), neural networks, frequency-domain bootstrap and kernel ridge regression\cite{Utama18,Niu2019,Neufcourt2018,Wu2020a,Yuksel2021,Gao2021,Shelley2021,Sharma2022,Navarro2022,Lovell2022,Mumpower2023} (see the recent review \cite{Boehnlein2022} for more references). 
In a series of papers \cite{Neufcourt2019,Neufcourt2020a, Neufcourt2020b, Kejzlar2020,Hamaker2021}, the BMA methodology has been applied to nuclear mass predictions. In this work, 
we propose the LBMM approach to produce
model-informed extrapolations of nuclear masses
that overcome the limits of BMA mentioned above.

\section*{Bayesian model mixing}\label{sec:LBMM}

Let us 
consider experimental observations $y(x_i)$ of a physical process at locations $x_i\in \mathcal{X} \subset \mathbb{R}^q$, $i=1,\ldots,n$,
governed by a ``true model'' $y^*(x)$,
and let us assume that the true model is not fully captured by one of the proposed models $f_k, k=1,\ldots,p$,
but rather a combination of these models. 
It is then natural to consider 
a statistical mixture model of the general form:
\begin{equation}
    \label{eqn:LMM-model}
    y(x_i) = \sum_{k=1}^p \omega_k(x_i) f_k(x_i) + \sigma \epsilon_i,
\end{equation}
where $\sigma$ represents the scale of the error of the mixture model, $\epsilon_i \overset{\mathrm{iid}}{\sim} N(0,1) $, and $f_1(x_i), \dots, f_p(x_i)$ are theoretical values for the datum  $y(x_i)$ provided by the $p$ theoretical models considered.

In practice, the weights $\omegab(x_i) = (\omega_1(x_i), \dots, \omega_p(x_i))$ must be taken in a space where inference is possible. 
This can be done in many ways. In this work, we will highlight a few alternative models for $\omegab(x_i)$  which are tractable, suggestive, and fully Bayesian. 

Additionally, one can improve the models by accounting for systematic errors. This can be done by adding to each model
the systematic error correction $\delta_{f,k}$:
\begin{equation}
\label{eqn:LMM-model_local}
    y(x_i) = \sum_{k=1}^p \omega_k(x_i) \left(f_k(x_i) +\delta_{f,k}(x_i)\right)
    + \sigma \epsilon_i.
\end{equation}

\subsection*{Global mixtures\label{sec:global_weights}}

First, we present the simplest application of Bayesian model mixing (BMM) where one assumes global weights (GBMM),
i.e., weights that are \emph{constant over the input domain}.

\subsubsection*{Linear model (L)}

Let us first model the underlying physical process by a global (linear) mixture of the individual models:
\begin{equation}\label{eqn:linear_model}
    y(x_i) = \sum_{k=1}^p \omega_k \left(f_k(x_i)+\delta_{f,k}(x_i)\right) + \sigma \epsilon_i.
\end{equation}
Using Eq.~(\ref{eqn:linear_model})
the log-likelihood of the model can be written as:
\begin{align}
&\log p(\yb|\omegab,\deltab_f,\sigma) 
=
-\frac{n}{2}\log(2\pi \sigma^2) \nonumber -\frac{1}{2\sigma^2}\sum_{i=1}^n  \left[y(x_i) - \sum_{k=1}^p \omega_k \left(f_k(x_i)+\delta_{f,k}(x_i)\right)\right]^2,
\end{align}
where $\yb = (y(x_1), \dots, y(x_n))$. To ensure that the weights $\omegab$ have the same support as the model weights in BMA, it may also be justified to assume the simplex constraints $\omega_k\geq 0$ and $\sum_k\omega_k=1$.
In that case, the posterior distributions should also satisfy the simplex constraints.
While the first condition is easily met using non-negative priors,
the second can be more challenging to enforce with priors.
Nevertheless the naive idea
of projecting the unconstrained posteriors 
appears to be relatively efficient in the case of simple linear models \cite{Patra2019}.
Here projecting the simplex constraints  
corresponds to substituting 
\begin{equation}
\label{eq:constraint}
\omega_k \longleftarrow \frac{max(\omega_k, 0)}{\sum_\ell\max(\omega_\ell, 0)}.
\end{equation}

\subsubsection*{Dirichlet model (D) \label{subsub:Dirichlet}}

\begin{figure}[htb]
\centering
		\includegraphics[width=.8\linewidth]{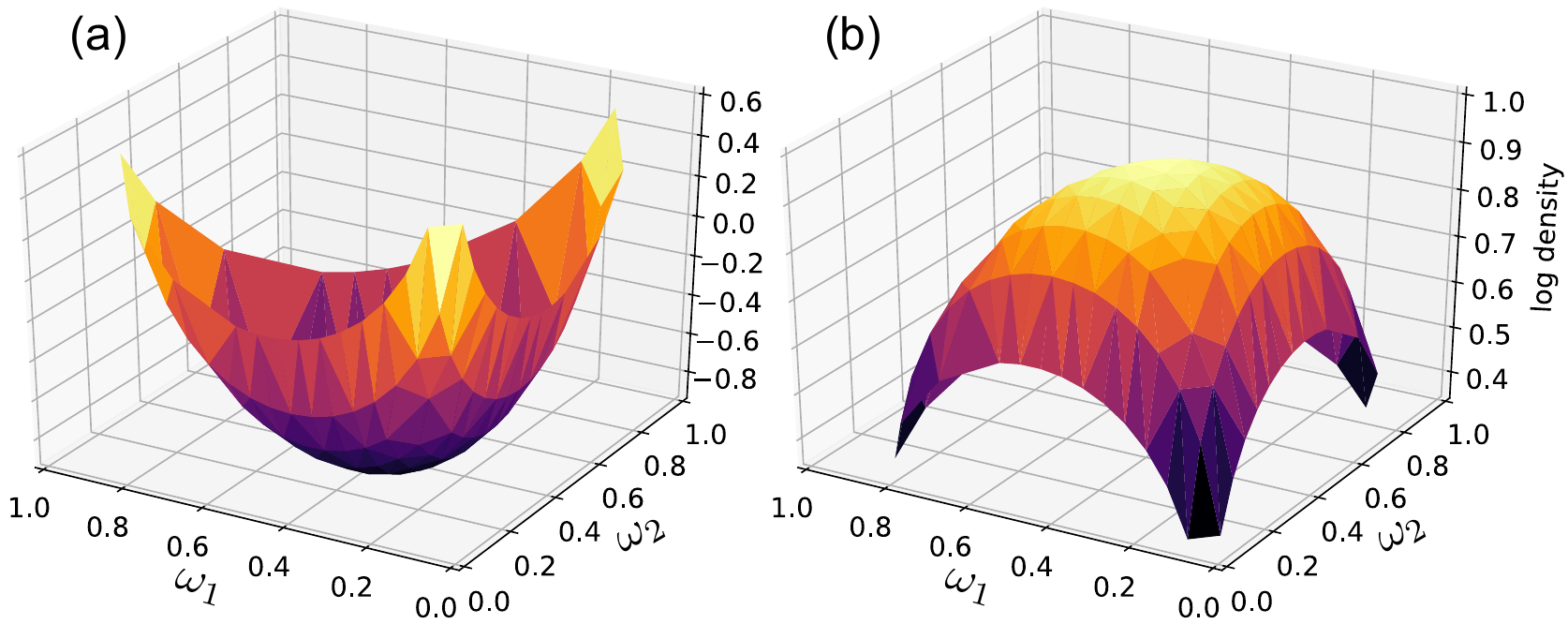}
		\caption{Log density of the Dirichlet distribution when $p =3$ as a function of $\omega_1$ and $\omega_2$ ($\omega_3=1-\omega_1-\omega_2$). The parametrization of Dirichlet distribution is such that $\alphab <1$ is close to model selection (one dominant weight with a high probability) and $\alphab > 1$ leads to true mixing 
  (several large-probability weights). Left:  $\alphab = (0.3,0.3,0.3)$; Right:  $\alphab = (1.3,1.3,1.3)$.\label{fig:dirichlet}}
\end{figure}

As a refinement of the global linear mixture and a step towards local mixtures with simplex constraints, one can suppose that the weights $\omegab$ are given \emph{hierarchically} by a Dirichlet distribution:
\begin{equation}
p(\omegab|\alphab) 
\propto \prod_{k} \omega_k^{\alpha_k - 1}
\end{equation}
with the hyperprior $\pi(\alphab)$ on the hyperparameter $\alphab$. The reason for this additional modeling layer is twofold. First, it allows us to express uncertainty about the prior model weighing imposed by $\alphab$. 
Note that a Dirichlet distribution with size $p$ 
and the parameter vector $\alphab > 0$
is a multivariate continuous distribution
on the simplex $\{\omega_1, ..., \omega_p \geq 0: \sum_{k=1}^p  \omega_k = 1\}$ 
where the average value of $\omega_j$ is $\alpha_j / \sum_{k=1}^p \alpha_k$.
Looking at the shape of the distribution in Fig. \ref{fig:dirichlet},
it is clear that  $\alphab < 1$ is close to model selection
while $\alphab > 1$ encourages true mixing. 
The hyperprior $\pi(\alphab)$ allows us to quantify our uncertainty about these two regimes. 
Secondly, the hierarchical model for $\omegab$ 
permits, with slight modification, a heterogeneity of weights based on the value of $x$, which we shall exploit shortly. 

Consequently, up to a choice of prior
$\pi(\alphab,\deltab_f,\sigma)$,
the joint posterior distribution of $(\omegab, \alphab, \deltab_f,\sigma)$ is given by 
\begin{equation}
p(\omegab, \alphab,\deltab_f,\sigma|\yb)
\propto
p(\yb|\omegab, \deltab_f,\sigma) 
p(\omegab|\alphab)
\pi(\alphab,\deltab_f,\sigma),
\end{equation}
which does not have a closed form in general and needs to be approximated using MCMC. Predictions of observations from a physical process at new locations are then obtained by propagating the posterior samples of $(\alphab, \deltab_f,\sigma)$ through the hierarchy described above. 

The Dirichlet weights encapsulate the contribution of each model to the mixture, which makes the interpretation of the weights probabilistic. In that sense, they carry a different meaning than
the BMA weights, which measure the fidelity of individual models.
Still, both weights can be compared as they play the same role in the the final predictions
-- keeping in mind that the information
contained in the posterior distribution of the Dirichlet weights
is richer than the point values produced by BMA.

\subsection*{Local mixtures}

Let us now consider that the observations $y(x_i)$
follow the general statistical model given by Eq.~\eqref{eqn:LMM-model_local}. The key feature here is that now the weights depend on the location $x$.
Without additional information,
the functions $\omegab(x_i)$ shall be estimated
with a non-parametric estimator
satisfying the 
simplex constraints.

In order to account for the local dependency of the model weights while satisfying the simplex constraints, we propose a hierarchical framework based on the Dirichlet distribution. Specifically, 
we take for every $x$ weights $\omegab(x)$ as Dirichlet random variables 
defined by parameters $\alphab(x) = (\alpha_1(x), \dots, \alpha_p(x))$.
The correlations between $\omega_k(x)$'s
for different values of $x$ shall 
be contained in the corresponding $\alpha_k(x)$-correlations. We investigate two models for $\alpha_k(x)$: a 
Generalized Linear Dirichlet model (GLD) and a Gaussian Process Dirichlet model (GPD). 

In particular, we assume that at every location $x$
the model weights $\omegab(x)$ follow
a Dirichlet distribution 
with parameters $\alpha_k (x)$, for $k = 1, \dots, p$, now depending on $x$, that will encode the spatial relationships between the model weights over the input space.
Since the Dirichlet distribution is defined 
for parameter values $\alpha_k>0$, we additionally apply an exponential \emph{link} function,
i.e., we consider 
$\gamma_k(x):=\log(\alpha_k(x))$
which can be regressed symmetrically.
Thus, the purpose of the link function is to allow for unconstrained modeling of $\gamma_k(x)$. 

The GLD version of our local mixing framework represents 
$\gamma_k(x)$ parametrically as
\begin{equation}
    \gamma_k(x) = \betab^T_k x, 
    \label{eq:dirichlet_GLM}
\end{equation}
where $(\betab_1, \dots, \betab_p) = (\beta_{1,1}, \dots, \beta_{1,q}, \dots, \beta_{p,1}, \dots, \beta_{p,q})$ is a parameter vector. The linear nature of Eq.~\eqref{eq:dirichlet_GLM} corresponds to the assumptions that the correlations between local weights have a relatively large spatial range.

As a finer version of LBMM, we propose a non-parametric GPD model for $\gamma_k(x)$ defined by a Gaussian process prior parametrized with a constant mean $\gamma_k^{\infty}$ and covariances $c_k(x, x')$ given by quadratic exponential kernels \cite{RasmussenWilliams}. Additionally, we assume that $\gamma_k(x)$ and $\gamma_j(x)$ are statistically independent for $k \ne j$. Note that the proposed hierarchical structure takes into account not only the relationship between $\omega_k(x)$'s for different values of $x$ (via $\alpha_k (x)$) but also the correlations between the weights of models at given spatial location $x$ (via Dirichlet distribution). This would not be possible if one choose to model $\omega_k(x)$'s directly, let's say with a GP over $\omega_k(x)$.

\section*{Application: nuclear mass extrapolation}\label{sec:application}

As a case study for the Bayesian model mixing framework described above, we consider the separation energies of atomic nuclei, which were the subject of our previous investigations \cite{Neufcourt2018,Neufcourt2019,Neufcourt2020a,Neufcourt2020b,Hamaker2021}. Our particular goal is to compare the following alternatives:
\begin{enumerate}
    \item[(i)] Raw models without statistical correction $\delta_f$ (/w $\delta_f$) vs. models corrected  with $\delta_j$ (w/ $\delta_f$);
    \item[(ii)] BMA vs. global BMM;
    \item[(iii)] Global BMM vs. local BMM.
\end{enumerate}

\begin{figure}[htb]
\centering
    \includegraphics[width=0.5\linewidth]{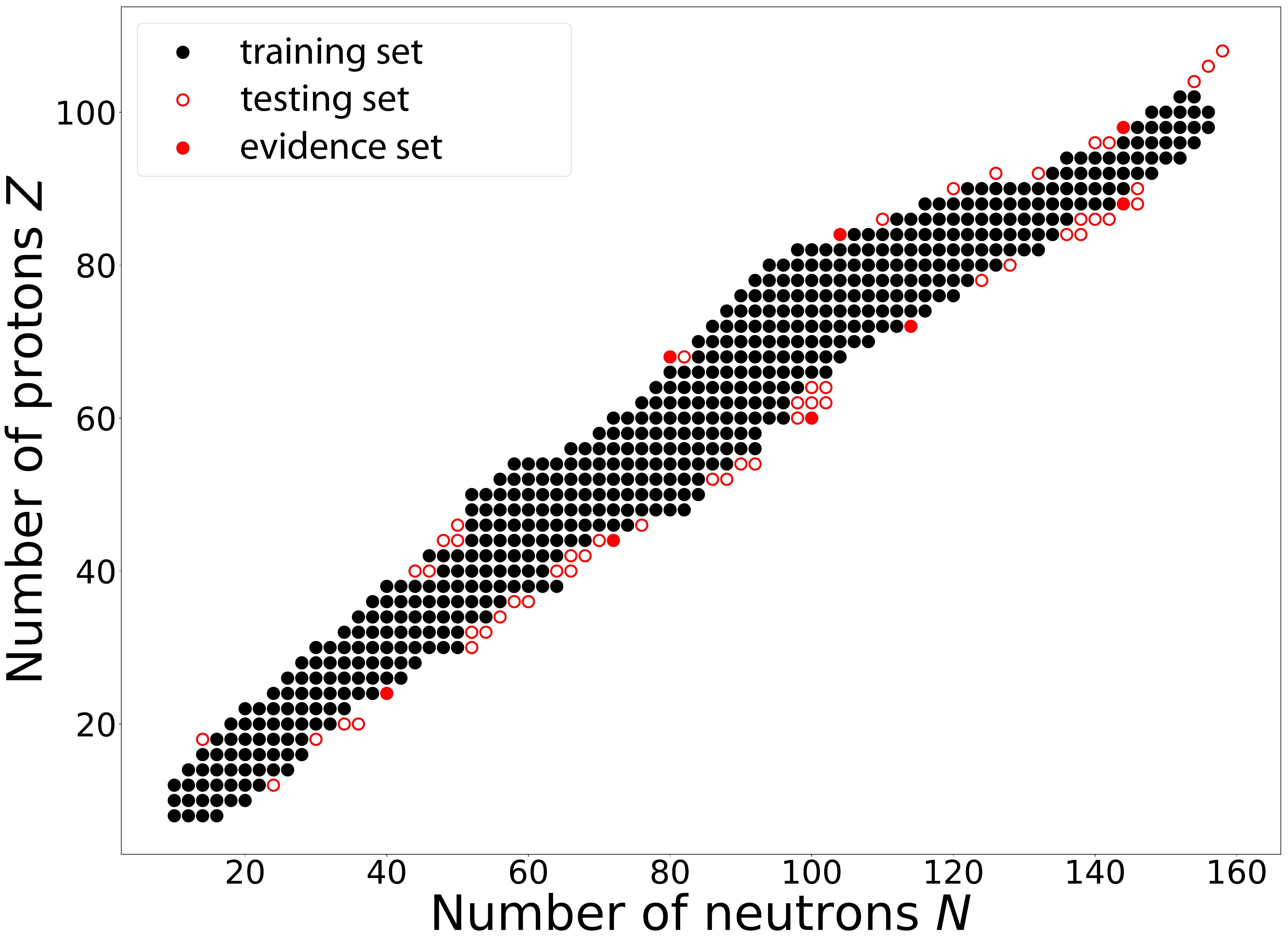}
    \caption{Training (black dots), testing (red circles),
    and evidence (red dots) datasets of two-neutron separation energies of even-even nuclei used in this study. The eight evidence nuclei are also included in the testing dataset. Each nucleus  is represented by the number of protons $Z$ and neutrons $N$. See text for details.}
    \label{fig:dataset}
\end{figure}

The two-neutron separation energy ($S_{2n}$) is a fundamental property of the atomic nucleus defined as the energy required to remove two neutrons
from the nucleus. It can be expressed as a difference of nuclear masses. 
Here, the input space $\mathcal{X}$ is represented by the numbers of protons $Z$ and neutrons $N$. Consequently, in this study $q=2$, $x_i := (Z_i, N_i)$
and $y_i$ is the observed two-neutron separation energy at $x_i$. We are particularly interested in even-even nuclei, for which both $N$ and $Z$ are even numbers. 
We use the most recent measured values of two-neutron separation energies for nuclei from the AME2003 dataset \cite{AME03b} as training data ($n = 521$)
for BMM and the GP systematic corrections; 
for BMA calculations we retain as evidence dataset a subset of this training data consisting of
8 nuclei: 3 proton-rich
nuclei $^{148}$Er, $^{188}$Po, $^{242}$Cf,  and 5 neutron-rich nuclei
$^{64}$Cr, $^{116}$Ru, $^{160}$Nd, $^{168}$Hf, $^{232}$Ra.
We keep additional data tabulated in AME2020 \cite{AME2020a} for an out of sample extrapolative testing dataset $(n=59)$. 
These three domains are depicted in Fig.~\ref{fig:dataset}.

As for prediction, we will use the largest domain on which two-neutron separation energies are positive, i.e.,
the corresponding nuclei are predicted to exist.
In line with our previous studies, 
we consider seven theoretical models based on the nuclear density functional theory (DFT) which is capable of   describing  
the whole nuclear chart: SkM$^*$ \cite{Bartel1982}, 
SkP \cite{Dob84}, 
SLy4 \cite{Chabanat1995}, 
SV-min \cite{Kluepfel2009}, 
UNEDF0 \cite{UNEDF0}, 
UNEDF1 \cite{UNEDF1}, and
UNEDF2 \cite{UNEDF2}. The DFT data were taken from the theoretical database \cite{massexplorer}.
The above set of DFT models was augmented by two well-fitted  mass  models FRDM-2012 \cite{Moller2012}
and HFB-24 \cite{Goriely2013}
that  have significantly more parameters than the (less phenomenological) DFT models, 
resulting in a
better fit to measured masses.

In the subsequent Bayesian analyses, we use independent priors for the different statistical parameters 
and keep consistent notation throughout 
the model mixing variants.
In general, we use normal priors for unconstrained parameters,
Gamma priors for positive parameters,
and uniform priors for bounded parameters.
Recall that a Gamma distribution is parametrized by
a shape parameter $a$ and a rate parameter $b$,
and has its mean given by ${a}/{b}$ and variance given by 
${a}/{b^2}$.
For the error scale parameters $\sigma$,
we use Gamma priors
with scale parameter 5 and rate parameter 10,
with mean 0.5\,MeV and standard deviation 0.22\,MeV.

In the case of LBMM variant with GLD defined by Eq.~\eqref{eq:dirichlet_GLM},
we take independent normal prior distributions with mean 0 and standard deviation 1 for the elements of 
$\betab_1, \dots, \betab_p$. For LBMM with GPD, we used
independent squared exponential kernels for the GP: namely,
$$c_k(x, x') = \eta_k 
e^{-\frac{(Z - Z')^2}{2\rho_{Z}^2}
-\frac{(N - N')^2}{2\rho_{N}^2}}$$ 
characterized by three hyperparameters  
$\eta_k, \rho_{Z},$ and $\rho_{N}$. We have chosen to take the length-scale parameters $\rho$ common to all nuclear models,
but leave the GP intensity parameters $\eta$ be different for each model, in order to ensure stability and convergence.
As a result the GPD weights follow the frequency of the residuals for each model. In the case of Dirichlet distribution with GPD local mixture, we take independent normal priors with mean 0 and variance 1 for the GP mean parameter $\gamma_k^\infty$,
and independent Gamma priors for the three scale parameters
$\eta_k$, $\rho_{Z}$ and $\rho_{N}$. 
These priors are taken with respective parameters
$(10, 2)$, $(5, 2)$, $(5, 2)$; 
this corresponds to slightly informative priors
which helped to ensure convergence towards weights 
localized on an appropriate scale. The parameter $\gamma_k^\infty$ determines the long range weight of each model,
i.e., far from the training data. 
Note that taking a zero-mean GP, i.e., setting $\gamma_k^\infty = 0$,
would amount to uniform weights far from the data. 

When it comes to the global GBMM+L mixture, Eq.~\eqref{eqn:LMM-model}, we take for 
$\omegab$ independent uniform priors on $[0, 1]$.
In practice, the simplex constraint
is satisfied implicitly, without the need to apply Eq.~\eqref{eq:constraint}.
This confirms that all the individual models are well conceived. For the GBMM+D variant, we take for $\alpha$ a half-normal prior with standard deviation 2. 

For $\delta_{f,k}$, we use the systematic correction for two-neutron separation energy residuals (i.e., differences between  theoretical and measured two-neutron separation energies) computed in \cite{Neufcourt2020b} using Bayesian Gaussian processes; these are fixed with no priors. The training dataset in \cite{Neufcourt2020b} agrees with Fig. \ref{fig:dataset} up to a set of 5 additional nuclei. In what follows, we do not use these nuclei during training whenever uncorrected models are considered since they would be extrapolative from the GP's perspective. The consequences of this omission is negligible due to the overall size of the training set.

All model weights were trained with the $S_{2n}$ values from the full training dataset with the exception of BMA, 
where we have used only the evidence set shown in Fig.~\ref{fig:dataset} that consists
of 8 nuclei. Similar to \cite{Neufcourt2019, Neufcourt2020a, Neufcourt2020b}, we compute the BMA weights only on a set of representative nuclei
because computing evidences on a large dataset inevitably
leads to model selection. This happens due to the exponential and multiplicative nature of Gaussian likelihood which punishes large deviations more than it favors good fits (see \cite{Kejzlar2020,BANDmanifesto} for details). 
These weights are in turn applied 
to obtain predictions
for $S_{2n}$.
The proton-rich limit of $\mathcal{X}$, determined by two-proton separation energies, was identified in the previous study \cite{Neufcourt2020a}.

Tables~\ref{tab:rms} 
and \ref{tab:weights}
summarize the results of our model variants 
and are discussed in the following paragraphs.

\begin{table}[htb]
    \caption{
Rms deviations (in MeV) for all individual models,  global (BMA, GBMM) 
and local (LBMM) mixtures. 
Values are provided
with and without systematic corrections. 
For corrected models, we show only the test rms as the train rms and $\sigma$ values are negligible after the GP fit.  For abbreviations of BMA variants, see text.
    \label{tab:rms}}
    
    \centering
    \fbox{%
    \begin{tabular}{l|ccc|c}
     & \multicolumn{3}{c|}{Uncorrected models} & \multicolumn{1}{c}{Corrected models} \\
        Model & Train & Test & $\sigma$ & Test \\[4pt] 
        SkM$^*$ & 1.19 & 1.14 & 1.19(4) & 0.66  \\  
        SkP & 0.84 & 0.74 & 0.83(3) & 0.64  \\
        SLy4 & 0.99 & 0.81 & 0.99(3) &  0.68 \\ 
        SV-min & 0.77 & 0.63 & 0.77(2) & 0.55  \\
        UNEDF0 & 0.77 & 0.63 & 0.77(2) & 0.61  \\
        UNEDF1 & 0.75 & 0.50 & 0.75(2)  & 0.48 \\ 
        UNEDF2 & 0.85 & 0.67 & 0.84(3) & 0.54 \\
        FRDM-2012 & 0.48 & 0.45 & 0.47(1) & 0.38 \\
        HFB-24 & 0.42 & 0.40 & 0.42(1) & 0.40 \\[4pt]
        BMA(ex)  & 0.38 & 0.32 & 0.55(16) & 0.35\\
        BMA(MC) & 0.39 & 0.32 & 0.56(17) &  0.35  \\
        BMA(Lap) & 0.40 & 0.32 & 0.57(17) &0.35\\[4pt]
        GBMM+L &  0.33 & 0.31 & 0.33(1) &  0.41 \\
        GBMM+D &  0.33 & 0.31 & 0.33(1) &  0.46  \\[4pt]
        LBMM+GLD & 0.29 & 0.35 & 0.30(1) \\
        LBMM+GPD & 0.25 & 0.33 & 0.26(1) \\
    \end{tabular}}
\end{table}

\begin{figure*}[ht!]
    \centering
    \includegraphics[width=0.8\linewidth]{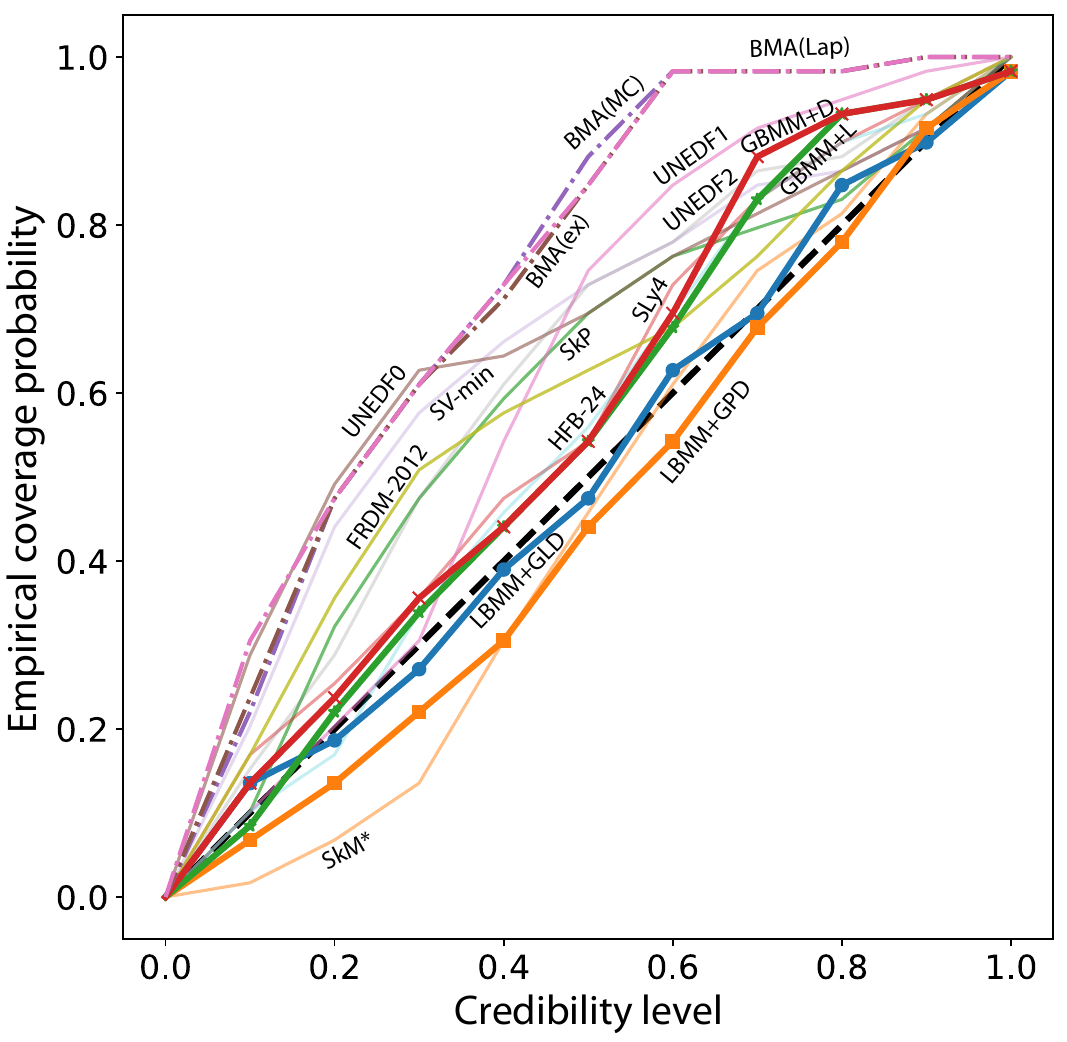}
    \caption{Empirical coverage probability for raw models without statistical correction together with BMA and BMM variants. The empirical coverage probability  was calculated with equal-tailed credibility intervals.
    The  reference line (diagonal) is marked by a dashed line. \label{fig:ECP}}
\end{figure*}

\subsection*{Uncorrected models versus corrected models}

We first discuss the fidelity of 
 individual models. To this end, we study root-mean-square (rms) deviations
for all the modeling variants discussed in this paper. We consider raw model predictions
and the predictions  including the systematic corrections $\delta_{f,k}$.
It is seen that for the individual models, the corrected variants generally  outperform the raw predictions, see Ref.~\cite{Neufcourt2018} for discussion. The exception is HFB-24, which has been carefully calibrated to experimental masses; in this case the  correction term 
$\delta_f$ 
does not lead to a lower rms deviation on the testing dataset.
 We want to point out that mixing of corrected models should be done with caution. It is our empirical experience that \emph{mixing (GBMM or LBMM) of previously corrected models can lead to overfitting} as one tends to fit the statistical models to the small leftover  noise since the residuals of all corrected models on the training dataset  are practically zero. This can be clearly observed in Table~\ref{tab:rms}: the rms  deviations for both local and global mixtures slightly outperform the combinations of corrected models on the the testing dataset. For instance, GBMM+D of uncorrected models gives 0.31 MeV rms deviation on the testing dataset as compared to 0.46 MeV on the corrected models and also as compared to 0.35 MeV rms deviations of BMA on corrected models. Since providing accurate extrapolations is the main focus of this work, in the following, we focus the discussions primarily on the uncorrected models.

\begin{table*}[ht!]
    \caption{    \label{tab:weights}
        Global weights calculated on the training dataset with different methods:
        BMA on the evidence subset
        (see Fig \ref{fig:dataset}) 
        obtained from a closed form computation
        as well as Monte Carlo and Laplace approximations,
        and the two global mixtures GBMM+L and GBMM+D obtained over the whole training set.
 For compactness, the following abbreviations are used: SV=SV-min, UNEn=UNEDFn (n=0,1,2),  FRDM=FRDM-2012, and HFB=HFB-24.}
    \centering
    \setlength{\tabcolsep}{3pt}
    \fbox{%
    \begin{tabular}{l|ccccccccccc}
        Model & SkM$^*$ & SkP & SLy4 & SV &
        UNE0 & UNE1 & UNE2 & FRDM & HFB  \\
        \hline
 & \multicolumn{9}{c}{Uncorrected models} \\
        BMA(ex) & 0.00 & 0.00 & 0.00 & 0.03 & 0.07 & 0.02 & 0.13 & 0.48 & 0.27 \\
        BMA(MC) & 0.00 & 0.00 & 0.00 & 0.04 & 0.09 & 0.04 & 0.15 & 0.42 & 0.26 \\
        BMA(Lap) & 0.00 & 0.00 & 0.00 & 0.03 & 0.09 & 0.04 & 0.16 & 0.46 & 0.22\\[4pt]
        GBMM+L &  0.01 & 0.02 & 0.01 & 0.01 & 0.02 & 0.02 & 0.06 & 0.28 & 0.57 \\
        GBMM+D & 0.01& 0.02 & 0.00 & 0.00 & 0.02 & 0.01 & 0.07 & 0.29 & 0.57\\
\hline \\[-5pt]
 & \multicolumn{9}{c}{Corrected models} \\
        BMA(ex) & 0.00 & 0.00 & 0.00 & 0.01 & 0.00 & 0.44 & 0.06 & 0.47 & 0.02 \\
        BMA(MC) & 0.00 & 0.00 & 0.00 & 0.02 & 0.01 & 0.41 & 0.09 & 0.43 & 0.03 \\
        BMA(Lap) & 0.00 & 0.00 & 0.00 & 0.02 & 0.01 & 0.42 & 0.09 & 0.44 & 0.02\\[4pt]
        GBMM+L &  0.11 & 0.12 & 0.11 & 0.13 & 0.08 & 0.11 & 0.13 & 0.10 & 0.11 \\
        GBMM+D & 0.24 & 0.15 & 0.20 & 0.13 & 0.03 & 0.02 & 0.10 & 0.00 & 0.12 \\
    \end{tabular}}
\end{table*}

\subsection*{BMA versus Global Mixtures}\label{sec:BMA:results}

The BMA evidence integrals were calculated 
on the evidence dataset
by means of Monte Carlo (MC), Laplace approximations, 
and in a closed form under conjugate priors. We denote the corresponding BMA variants as follows: 
BMA(MC), BMA(Lap) and BMA(ex), respectively
(see Methods section for the calculations of BMA weights). 
In Table~\ref{tab:weights}, we see that the model weights 
produced by BMA 
are consistent across all three evidence computation approaches, irrespective of whether the systematic correction has been applied. Averaging corrected models is more democratic as compared to the raw models: this is expected since the GP-based $\delta_{f,k}$ corrections fit the training data closely irrespective of the theoretical model. Still, the BMA testing rms with uncorrected and corrected models are very similar, with a slight preference for the uncorrected models.

The global mixtures of uncorrected  models are generally slightly outperforming BMA on both training and testing datasets (see Table~\ref{tab:rms}).
This is in fact expected,
given that these weights are designed to maximize the predictive power of the model mixture.
Indeed, the GBMM+L model is the Bayesian counterpart to a 
frequentist linear regression against the 
different nuclear model predictions
that minimizes the 
rms on the training set.
This principle still holds despite the uniform prior used for the GBMM+L model
which is very informative 
and plays a regularizing role 
that reduces overfitting and favors mixing. 
We can see that the Dirichlet mixture model yields very similar weights, 
with the benefits of having its weights
natively located on the simplex.
This comparison of global weights already speaks in favour of ruling out BMA for the purpose of combining imperfect models, in the favor of a Bayesian Dirichlet model. 
Table~\ref{tab:rms} also shows the posterior
mean of the noise scale parameter $\sigma$ for comparison.
As a rule of thumb, a statistical model with a conservative (liberal) uncertainty quantification (UQ) would have $\sigma$ above (below) the test rms, and a statistical model with high-fidelity UQ has $\sigma$ close to the test rms. A more comprehensive view of UQ that reflects the fully propagated prediction uncertainty can be gleaned from Fig.~\ref{fig:ECP} that shows the empirical coverage probability \cite{Gneiting2007,Raftery2007} (ECP). Each curve in Fig.~\ref{fig:ECP} corresponds to the proportion of predictions in the testing dataset falling into the respective credible intervals (equal-tailed credible intervals). If the ECP curve  closely follows the diagonal, then the actual fidelity of the credible interval corresponds to the nominal value. Thus we see that the \emph{
GBMM
has both a superior prediction performance and a better UQ then BMA and individual models.}

\subsection*{Global versus Local Mixtures}

The  posterior mean of the LBMM+GPD weights 
are shown in Fig~\ref{fig:local_weights_gp}. 
The same plots but for LBMM+GLD are given in the Supplementary Information. As discussed earlier, mixing models locally corrected for systematic errors is highly susceptible to overfitting and we therefore focus on uncorrected models, i.e. without $\delta_{f}$.  Both LBMM variants show the dominance of the well-fitted HFB-24 mass model throughout the nuclear landscape. As expected, the simplistic linear dependence of weights $\omegab$ on $(Z,N)$ in the GLD variant is insufficient to fully capture the complex local behaviour of the mass models learned by a more flexible GPD variant.  
While the HFB-24 mass model dominates, the final LBMM+GPD results
involve other models, primarily FRDM2012, UNEDF0, and SkM$^*$.
The weight distribution naturally depends on the choice of models involved in the analysis. This suggests, that a preselection of diversified models to be used in LBMM could also be considered beforehand.

\begin{figure*}[htb]
    \centering
    \includegraphics[width=1.\linewidth]{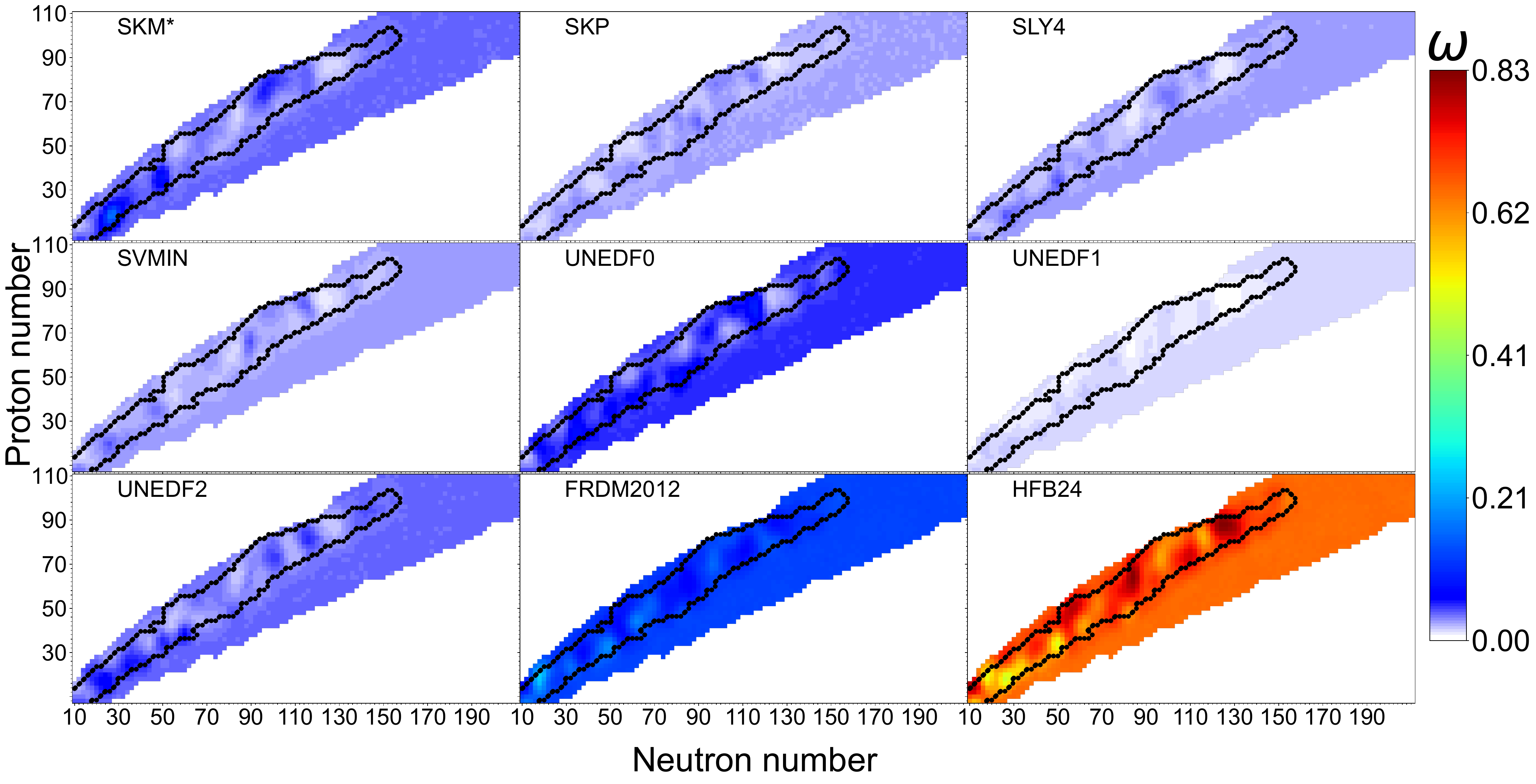}
    \caption{Posterior means of the local model weights in the LBMM+GPD variant across the nuclear landscape. \label{fig:local_weights_gp}}
\end{figure*}

In terms of the rms deviations, 
the GPD variant does better than the GLD local mixture, which reflects the ability of the GPD 
to capture the local performance of mass models. 
Local mixtures perform better
than global mixtures 
on the training set, 
and than BMA on both training and testing sets;
however they fall slightly behind global mixtures 
on the testing set.
We attribute this to the difficult tuning of the statistical model which is very sensitive to the variations of the parameters and a limited testing dataset in terms of its distribution across the nuclear landscape. 
In fact, Markov chain Monte Carlo (MCMC) sampling from the Bayesian posterior distributions can be numerically unsatisfactory with conventional Metropolis samplers. This is due to the relative large number of parameters of LBMM. In order to achieve satisfactory convergence, we recommend using more sophisticated No-U-Turn sampler \cite{NUTS} that tends to perform well in scenarios with moderately large parameter spaces (see Methods section for more details). In terms of UQ fidelity, the LBMM variants clearly dominate over the global mixtures, BMA, and individual models since the ECP of their respective predictive credible intervals closely matches the nominal values (see Fig.~\ref{fig:ECP}).

\section*{Discussion}
In this work, we propose and implement a Bayesian Dirichlet model mixing framework. The proposed method is illustrated by applying it to nuclear mass models to assess their local fidelity and improve  predictability. Raw theoretical models and their statistically-corrected versions were considered to better understand the interplay between GP modeling, BMA, and the  BMM frameworks.

Bayesian model mixing of raw models results in testing rms that are at least as good or better than Bayesian model averaging (irrespective of models being corrected) with clearly superior UQ. Thus, \emph{improving model predictions through mixing rather than mixing of corrected models} leads to the best performance in terms of both prediction accuracy and UQ. Since BMM is trained on a sizable training set, it is also more robust to the choice of priors than BMA which can be very prior sensitive if the evidence in data is weak \cite{Hoeting1999}. 

BMM of corrected models should be performed with caution as it may lead to overfitting. In this case, one likely achieves a better improvement with standard BMA based on a well chosen evidence set. 

The LBMM+GPD variant achieved the smallest training error on the training dataset (0.25 MeV) which demonstrates that the LBMM well captures the local presences of individual models. Furthermore, the local mixtures clearly surpass all the other modeling strategies explored in this work in terms of UQ fidelity. This shows that the proposed hierarchical Dirichlet model for LBMM effectively represents and propagates uncertainties which is essential for mass modeling into unexplored domains\cite{Neufcourt2020b}.

The results of BMA depend on the choice of the evidence dataset. That is, by increasing the density of the evidence data in the region of interest, e.g., for applications or extrapolations, one can improve the predictive power of averaging procedure.
Improvement in the performance of BMM can also be achieved by restricting the training dataset to the region of interest as opposed to training on the whole domain; 
this motivates our introduction of local BMM models. 

The distributions of BMA and BMM weights also depend on the choice of theoretical models. Table~\ref{tab:weights} and Fig.~\ref{fig:local_weights_gp} show that mixing a large set of models results in some having minimal contributions and point out to the existence of a class of models with similar local preferences (e.g., UNEDFn class). This indicates that adding model preselection and model orhogonalization \cite{Clyde1996} to the BMM pipeline could lead to a further improvement in predictive performance. 
In fact, in the context of our GBMM+L model,
it is well known that collinearity between the proposed theoretical models is a source of major 
instabilities.

\bibliographystylemethods{naturemag}
\bibliography{biblio}

\begin{thebibliography}{10}
\urlstyle{rm}
\expandafter\ifx\csname url\endcsname\relax
  \def\url#1{\texttt{#1}}\fi
\expandafter\ifx\csname urlprefix\endcsname\relax\def\urlprefix{URL }\fi
\expandafter\ifx\csname doiprefix\endcsname\relax\def\doiprefix{DOI: }\fi
\providecommand{\bibinfo}[2]{#2}
\providecommand{\eprint}[2][]{\url{#2}}

\bibitem{Hoeting1999}
\bibinfo{author}{Hoeting, J.~A.}, \bibinfo{author}{Madigan, D.},
  \bibinfo{author}{Raftery, A.~E.} \& \bibinfo{author}{Volinsky, C.~T.}
\newblock \bibinfo{journal}{\bibinfo{title}{{Bayesian} model averaging: a
  tutorial (with comments by {M}. {Clyde}, {David} {Draper} and {E}. {I}.
  {G}eorge, and a rejoinder by the authors}}.
\newblock {\emph{\JournalTitle{Statist. Sci.}}} \textbf{\bibinfo{volume}{14}},
  \bibinfo{pages}{382--417}, \doiprefix\url{10.1214/ss/1009212519}
  (\bibinfo{year}{1999}).

\bibitem{Was00}
\bibinfo{author}{Wasserman, L.}
\newblock \bibinfo{journal}{\bibinfo{title}{{Bayesian} model selection and
  model averaging}}.
\newblock {\emph{\JournalTitle{J. Math. Psych.}}}
  \textbf{\bibinfo{volume}{44}}, \bibinfo{pages}{92--107},
  \doiprefix\url{10.1006/jmps.1999.1278} (\bibinfo{year}{2000}).

\bibitem{Fragoso2018}
\bibinfo{author}{Fragoso, T.}, \bibinfo{author}{Bertoli, W.} \&
  \bibinfo{author}{Louzada, F.}
\newblock \bibinfo{journal}{\bibinfo{title}{Bayesian model averaging: A
  systematic review and conceptual classification}}.
\newblock {\emph{\JournalTitle{Int. Stat. Rev.}}}
  \textbf{\bibinfo{volume}{86}}, \bibinfo{pages}{1--28},
  \doiprefix\url{10.1111/insr.12243} (\bibinfo{year}{2018}).

\bibitem{BANDmanifesto}
\bibinfo{author}{Phillips, D.~R.} \emph{et~al.}
\newblock \bibinfo{journal}{\bibinfo{title}{Get on the {BAND wagon: A Bayesian}
  framework for quantifying model uncertainties in nuclear dynamics}}.
\newblock {\emph{\JournalTitle{J. Phys. G}}} \textbf{\bibinfo{volume}{48}},
  \bibinfo{pages}{072001}, \doiprefix\url{10.1088/1361-6471/abf1df}
  (\bibinfo{year}{2021}).

\bibitem{MWR2011}
\bibinfo{author}{Chmielecki, R.~M.} \& \bibinfo{author}{Raftery, A.~E.}
\newblock \bibinfo{journal}{\bibinfo{title}{Probabilistic visibility
  forecasting using {Bayesian} model averaging}}.
\newblock {\emph{\JournalTitle{Mon. Wea. Rev.}}}
  \textbf{\bibinfo{volume}{139}}, \bibinfo{pages}{1626--1636},
  \doiprefix\url{10.1175/2010MWR3516.1} (\bibinfo{year}{2011}).

\bibitem{Montgomery2010}
\bibinfo{author}{Montgomery, J.~M.} \& \bibinfo{author}{Nyhan, B.}
\newblock \bibinfo{journal}{\bibinfo{title}{Bayesian model averaging:
  {Theoretical} developments and practical applications}}.
\newblock {\emph{\JournalTitle{Political Anal.}}}
  \textbf{\bibinfo{volume}{18}}, \bibinfo{pages}{245--270},
  \doiprefix\url{10.1093/pan/mpq001} (\bibinfo{year}{2010}).

\bibitem{Zou2012}
\bibinfo{author}{Zou, Y.}, \bibinfo{author}{Lord, D.}, \bibinfo{author}{Zhang,
  Y.} \& \bibinfo{author}{Peng, Y.}
\newblock \bibinfo{journal}{\bibinfo{title}{Application of the {Bayesian} model
  averaging in predicting motor vehicle crashes}}.
\newblock {\emph{\JournalTitle{US Department of Transportation}}}
  (\bibinfo{year}{2012}).

\bibitem{Neufcourt2019}
\bibinfo{author}{Neufcourt, L.}, \bibinfo{author}{Cao, Y.},
  \bibinfo{author}{Nazarewicz, W.}, \bibinfo{author}{Olsen, E.} \&
  \bibinfo{author}{Viens, F.}
\newblock \bibinfo{journal}{\bibinfo{title}{Neutron drip line in the {Ca}
  region from {Bayesian Model Averaging}}}.
\newblock {\emph{\JournalTitle{Phys. Rev. Lett.}}}
  \textbf{\bibinfo{volume}{122}}, \bibinfo{pages}{062502},
  \doiprefix\url{10.1103/PhysRevLett.122.062502} (\bibinfo{year}{2019}).

\bibitem{Everett2021}
\bibinfo{author}{Everett, D.} \emph{et~al.}
\newblock \bibinfo{journal}{\bibinfo{title}{Phenomenological constraints on the
  transport properties of qcd matter with data-driven model averaging}}.
\newblock {\emph{\JournalTitle{Phys. Rev. Lett.}}}
  \textbf{\bibinfo{volume}{126}}, \bibinfo{pages}{242301},
  \doiprefix\url{10.1103/PhysRevLett.126.242301} (\bibinfo{year}{2021}).

\bibitem{Clemen1989}
\bibinfo{author}{Clemen, R.~T.}
\newblock \bibinfo{journal}{\bibinfo{title}{Combining forecasts: A review and
  annotated bibliography}}.
\newblock {\emph{\JournalTitle{Int. J. Forecast.}}}
  \textbf{\bibinfo{volume}{5}}, \bibinfo{pages}{559--583},
  \doiprefix\url{10.1016/0169-2070(89)90012-5} (\bibinfo{year}{1989}).

\bibitem{LeClarke2017}
\bibinfo{author}{Le, T.} \& \bibinfo{author}{Clarke, B.}
\newblock \bibinfo{journal}{\bibinfo{title}{{A Bayes Interpretation of Stacking
  for $\mathcal{M}$-Complete and $\mathcal{M}$-Open Settings}}}.
\newblock {\emph{\JournalTitle{Bayesian Anal}}} \textbf{\bibinfo{volume}{12}},
  \bibinfo{pages}{807 -- 829}, \doiprefix\url{10.1214/16-BA1023}
  (\bibinfo{year}{2017}).

\bibitem{StackingGelman18}
\bibinfo{author}{Yao, Y.}, \bibinfo{author}{Vehtari, A.},
  \bibinfo{author}{Simpson, D.} \& \bibinfo{author}{Gelman, A.}
\newblock \bibinfo{journal}{\bibinfo{title}{{Using Stacking to Average Bayesian
  Predictive Distributions (with Discussion)}}}.
\newblock {\emph{\JournalTitle{Bayesian Anal.}}} \textbf{\bibinfo{volume}{13}},
  \bibinfo{pages}{917 -- 1007}, \doiprefix\url{10.1214/17-BA1091}
  (\bibinfo{year}{2018}).

\bibitem{StackingGelman22}
\bibinfo{author}{Yao, Y.}, \bibinfo{author}{Pirš, G.},
  \bibinfo{author}{Vehtari, A.} \& \bibinfo{author}{Gelman, A.}
\newblock \bibinfo{journal}{\bibinfo{title}{{Bayesian Hierarchical Stacking:
  Some Models Are (Somewhere) Useful}}}.
\newblock {\emph{\JournalTitle{Bayesian Anal.}}} \textbf{\bibinfo{volume}{17}},
  \bibinfo{pages}{1043 -- 1071}, \doiprefix\url{10.1214/21-BA1287}
  (\bibinfo{year}{2022}).

\bibitem{Semposki22}
\bibinfo{author}{Semposki, A.~C.}, \bibinfo{author}{Furnstahl, R.~J.} \&
  \bibinfo{author}{Phillips, D.~R.}
\newblock \bibinfo{journal}{\bibinfo{title}{Interpolating between small- and
  large-$g$ expansions using {Bayesian} model mixing}}.
\newblock {\emph{\JournalTitle{Phys. Rev. C}}} \textbf{\bibinfo{volume}{106}},
  \bibinfo{pages}{044002}, \doiprefix\url{10.1103/PhysRevC.106.044002}
  (\bibinfo{year}{2022}).

\bibitem{yannotty2023model}
\bibinfo{author}{Yannotty, J.~C.}, \bibinfo{author}{Santner, T.~J.},
  \bibinfo{author}{Furnstahl, R.~J.} \& \bibinfo{author}{Pratola, M.~T.}
\newblock \bibinfo{title}{Model mixing using {Bayesian} additive regression
  trees} (\bibinfo{year}{2023}).
\newblock \eprint{2301.02296}.

\bibitem{Utama18}
\bibinfo{author}{Utama, R.} \& \bibinfo{author}{Piekarewicz, J.}
\newblock \bibinfo{journal}{\bibinfo{title}{Validating neural-network
  refinements of nuclear mass models}}.
\newblock {\emph{\JournalTitle{Phys. Rev. C}}} \textbf{\bibinfo{volume}{97}},
  \bibinfo{pages}{014306}, \doiprefix\url{10.1103/PhysRevC.97.014306}
  (\bibinfo{year}{2018}).

\bibitem{Niu2019}
\bibinfo{author}{Niu, Z.~M.}, \bibinfo{author}{Fang, J.~Y.} \&
  \bibinfo{author}{Niu, Y.~F.}
\newblock \bibinfo{journal}{\bibinfo{title}{Comparative study of radial basis
  function and {Bayesian} neural network approaches in nuclear mass
  predictions}}.
\newblock {\emph{\JournalTitle{Phys. Rev. C}}} \textbf{\bibinfo{volume}{100}},
  \bibinfo{pages}{054311}, \doiprefix\url{10.1103/PhysRevC.100.054311}
  (\bibinfo{year}{2019}).

\bibitem{Neufcourt2018}
\bibinfo{author}{Neufcourt, L.}, \bibinfo{author}{Cao, Y.},
  \bibinfo{author}{Nazarewicz, W.} \& \bibinfo{author}{Viens, F.}
\newblock \bibinfo{journal}{\bibinfo{title}{Bayesian approach to model-based
  extrapolation of nuclear observables}}.
\newblock {\emph{\JournalTitle{Phys. Rev. C}}} \textbf{\bibinfo{volume}{98}},
  \bibinfo{pages}{034318}, \doiprefix\url{10.1103/PhysRevC.98.034318}
  (\bibinfo{year}{2018}).

\bibitem{Wu2020a}
\bibinfo{author}{Wu, X.~H.} \& \bibinfo{author}{Zhao, P.~W.}
\newblock \bibinfo{journal}{\bibinfo{title}{Predicting nuclear masses with the
  kernel ridge regression}}.
\newblock {\emph{\JournalTitle{Phys. Rev. C}}} \textbf{\bibinfo{volume}{101}},
  \bibinfo{pages}{051301}, \doiprefix\url{10.1103/PhysRevC.101.051301}
  (\bibinfo{year}{2020}).

\bibitem{Yuksel2021}
\bibinfo{author}{Yüksel, E.}, \bibinfo{author}{Soydaner, D.} \&
  \bibinfo{author}{Bahtiyar, H.}
\newblock \bibinfo{journal}{\bibinfo{title}{Nuclear binding energy predictions
  using neural networks: {Application} of the multilayer perceptron}}.
\newblock {\emph{\JournalTitle{Int. J. Mod. Phys. E}}}
  \textbf{\bibinfo{volume}{30}}, \bibinfo{pages}{2150017},
  \doiprefix\url{10.1142/S0218301321500178} (\bibinfo{year}{2021}).

\bibitem{Gao2021}
\bibinfo{author}{Gao, Z.-P.} \emph{et~al.}
\newblock \bibinfo{journal}{\bibinfo{title}{Machine learning the nuclear
  mass}}.
\newblock {\emph{\JournalTitle{Nucl. Sci. Tech.}}}
  \textbf{\bibinfo{volume}{32}}, \bibinfo{pages}{109},
  \doiprefix\url{10.1007/s41365-021-00956-1} (\bibinfo{year}{2021}).

\bibitem{Shelley2021}
\bibinfo{author}{Shelley, M.} \& \bibinfo{author}{Pastore, A.}
\newblock \bibinfo{journal}{\bibinfo{title}{A new mass model for nuclear
  astrophysics: {Crossing 200 keV} accuracy}}.
\newblock {\emph{\JournalTitle{Universe}}} \textbf{\bibinfo{volume}{7}},
  \doiprefix\url{10.3390/universe7050131} (\bibinfo{year}{2021}).

\bibitem{Sharma2022}
\bibinfo{author}{Sharma, A.}, \bibinfo{author}{Gandhi, A.} \&
  \bibinfo{author}{Kumar, A.}
\newblock \bibinfo{journal}{\bibinfo{title}{Learning correlations in nuclear
  masses using neural networks}}.
\newblock {\emph{\JournalTitle{Phys. Rev. C}}} \textbf{\bibinfo{volume}{105}},
  \bibinfo{pages}{L031306}, \doiprefix\url{10.1103/PhysRevC.105.L031306}
  (\bibinfo{year}{2022}).

\bibitem{Navarro2022}
\bibinfo{author}{P{\'{e}}rez, R.~N.} \& \bibinfo{author}{Schunck, N.}
\newblock \bibinfo{journal}{\bibinfo{title}{Controlling extrapolations of
  nuclear properties with feature selection}}.
\newblock {\emph{\JournalTitle{Phys. Lett. B}}} \textbf{\bibinfo{volume}{833}},
  \bibinfo{pages}{137336}, \doiprefix\url{10.1016/j.physletb.2022.137336}
  (\bibinfo{year}{2022}).

\bibitem{Lovell2022}
\bibinfo{author}{Lovell, A.~E.}, \bibinfo{author}{Mohan, A.~T.},
  \bibinfo{author}{Sprouse, T.~M.} \& \bibinfo{author}{Mumpower, M.~R.}
\newblock \bibinfo{journal}{\bibinfo{title}{Nuclear masses learned from a
  probabilistic neural network}}.
\newblock {\emph{\JournalTitle{Phys. Rev. C}}} \textbf{\bibinfo{volume}{106}},
  \bibinfo{pages}{014305}, \doiprefix\url{10.1103/PhysRevC.106.014305}
  (\bibinfo{year}{2022}).

\bibitem{Mumpower2023}
\bibinfo{author}{Mumpower, M.} \emph{et~al.}
\newblock \bibinfo{journal}{\bibinfo{title}{Bayesian averaging for ground state
  masses of atomic nuclei in a machine learning approach}}.
\newblock {\emph{\JournalTitle{Front.Phys.}}} \textbf{\bibinfo{volume}{11}},
  \doiprefix\url{10.3389/fphy.2023.1198572} (\bibinfo{year}{2023}).

\bibitem{Boehnlein2022}
\bibinfo{author}{Boehnlein, A.} \emph{et~al.}
\newblock \bibinfo{journal}{\bibinfo{title}{Colloquium: Machine learning in
  nuclear physics}}.
\newblock {\emph{\JournalTitle{Rev. Mod. Phys.}}}
  \textbf{\bibinfo{volume}{94}}, \bibinfo{pages}{031003},
  \doiprefix\url{10.1103/RevModPhys.94.031003} (\bibinfo{year}{2022}).

\bibitem{Neufcourt2020a}
\bibinfo{author}{Neufcourt, L.} \emph{et~al.}
\newblock \bibinfo{journal}{\bibinfo{title}{Beyond the proton drip line:
  Bayesian analysis of proton-emitting nuclei}}.
\newblock {\emph{\JournalTitle{Phys. Rev. C}}} \textbf{\bibinfo{volume}{101}},
  \bibinfo{pages}{014319}, \doiprefix\url{10.1103/PhysRevC.101.014319}
  (\bibinfo{year}{2020}).

\bibitem{Neufcourt2020b}
\bibinfo{author}{Neufcourt, L.} \emph{et~al.}
\newblock \bibinfo{journal}{\bibinfo{title}{Quantified limits of the nuclear
  landscape}}.
\newblock {\emph{\JournalTitle{Phys. Rev. C}}} \textbf{\bibinfo{volume}{101}},
  \bibinfo{pages}{044307}, \doiprefix\url{10.1103/PhysRevC.101.044307}
  (\bibinfo{year}{2020}).

\bibitem{Kejzlar2020}
\bibinfo{author}{Kejzlar, V.}, \bibinfo{author}{Neufcourt, L.},
  \bibinfo{author}{Nazarewicz, W.} \& \bibinfo{author}{Reinhard, P.-G.}
\newblock \bibinfo{journal}{\bibinfo{title}{Statistical aspects of nuclear mass
  models}}.
\newblock {\emph{\JournalTitle{J. Phys. G}}} \textbf{\bibinfo{volume}{47}},
  \bibinfo{pages}{094001}, \doiprefix\url{10.1088/1361-6471/ab907c}
  (\bibinfo{year}{2020}).

\bibitem{Hamaker2021}
\bibinfo{author}{Hamaker, A.} \emph{et~al.}
\newblock \bibinfo{journal}{\bibinfo{title}{Precision mass measurement of
  lightweight self-conjugate nucleus $^{80}${Zr}}}.
\newblock {\emph{\JournalTitle{Nat. Phys.}}} \textbf{\bibinfo{volume}{17}},
  \bibinfo{pages}{1408--1412}, \doiprefix\url{10.1038/s41567-021-01395-w}
  (\bibinfo{year}{2021}).

\bibitem{Patra2019}
\bibinfo{author}{Patra, S.}
\newblock \emph{\bibinfo{title}{Constrained Bayesian Inference through
  Posterior Projection with Applications}}.
\newblock Ph.D. thesis (\bibinfo{year}{2019}).

\bibitem{RasmussenWilliams}
\bibinfo{author}{Rasmussen, C.~E.} \& \bibinfo{author}{Williams, C. K.~I.}
\newblock \emph{\bibinfo{title}{Gaussian Processes for Machine Learning}}
  (\bibinfo{publisher}{MIT Press}, \bibinfo{year}{2006}).

\bibitem{AME03b}
\bibinfo{author}{Audi, G.}, \bibinfo{author}{Wapstra, A.} \&
  \bibinfo{author}{Thibault, C.}
\newblock \bibinfo{journal}{\bibinfo{title}{{The {AME2003} atomic mass
  evaluation: (II). Tables, graphs and references}}}.
\newblock {\emph{\JournalTitle{Nucl. Phys. A}}} \textbf{\bibinfo{volume}{729}},
  \bibinfo{pages}{337 -- 676}, \doiprefix\url{10.1016/j.nuclphysa.2003.11.003}
  (\bibinfo{year}{2003}).

\bibitem{AME2020a}
\bibinfo{author}{Wang, M.}, \bibinfo{author}{Huang, W.},
  \bibinfo{author}{Kondev, F.}, \bibinfo{author}{Audi, G.} \&
  \bibinfo{author}{Naimi, S.}
\newblock \bibinfo{journal}{\bibinfo{title}{The {AME} 2020 atomic mass
  evaluation {(II). Tables,} graphs and references}}.
\newblock {\emph{\JournalTitle{Chin. Phys. C}}} \textbf{\bibinfo{volume}{45}},
  \bibinfo{pages}{030003}, \doiprefix\url{10.1088/1674-1137/abddaf}
  (\bibinfo{year}{2021}).

\bibitem{Bartel1982}
\bibinfo{author}{Bartel, J.}, \bibinfo{author}{Quentin, P.},
  \bibinfo{author}{Brack, M.}, \bibinfo{author}{Guet, C.} \&
  \bibinfo{author}{H{\aa}kansson, H.-B.}
\newblock \bibinfo{journal}{\bibinfo{title}{Towards a better parametrisation of
  {Skyrme-like} effective forces: A critical study of the {SkM} force}}.
\newblock {\emph{\JournalTitle{Nucl. Phys. A}}} \textbf{\bibinfo{volume}{386}},
  \bibinfo{pages}{79 -- 100}, \doiprefix\url{10.1016/0375-9474(82)90403-1}
  (\bibinfo{year}{1982}).

\bibitem{Dob84}
\bibinfo{author}{Dobaczewski, J.}, \bibinfo{author}{Flocard, H.} \&
  \bibinfo{author}{Treiner, J.}
\newblock \bibinfo{journal}{\bibinfo{title}{{Hartree-Fock-Bogolyubov}
  description of nuclei near the neutron-drip line}}.
\newblock {\emph{\JournalTitle{Nucl. Phys. A}}} \textbf{\bibinfo{volume}{422}},
  \bibinfo{pages}{103 -- 139}, \doiprefix\url{10.1016/0375-9474(84)90433-0}
  (\bibinfo{year}{1984}).

\bibitem{Chabanat1995}
\bibinfo{author}{Chabanat, E.}, \bibinfo{author}{Bonche, P.},
  \bibinfo{author}{Haensel, P.}, \bibinfo{author}{Meyer, J.} \&
  \bibinfo{author}{Schaeffer, R.}
\newblock \bibinfo{journal}{\bibinfo{title}{New {Skyrme} effective forces for
  supernovae and neutron rich nuclei}}.
\newblock {\emph{\JournalTitle{Physica Scr.}}} \textbf{\bibinfo{volume}{1995}},
  \bibinfo{pages}{231} (\bibinfo{year}{1995}).

\bibitem{Kluepfel2009}
\bibinfo{author}{Kl{\"{u}}pfel, P.}, \bibinfo{author}{Reinhard, P.-G.},
  \bibinfo{author}{B{\"{u}}rvenich, T.~J.} \& \bibinfo{author}{Maruhn, J.~A.}
\newblock \bibinfo{journal}{\bibinfo{title}{Variations on a theme by {Skyrme}:
  A systematic study of adjustments of model parameters}}.
\newblock {\emph{\JournalTitle{Phys. Rev. C}}} \textbf{\bibinfo{volume}{79}},
  \bibinfo{pages}{034310}, \doiprefix\url{10.1103/PhysRevC.79.034310}
  (\bibinfo{year}{2009}).

\bibitem{UNEDF0}
\bibinfo{author}{Kortelainen, M.} \emph{et~al.}
\newblock \bibinfo{journal}{\bibinfo{title}{Nuclear energy density
  optimization}}.
\newblock {\emph{\JournalTitle{Phys. Rev. C}}} \textbf{\bibinfo{volume}{82}},
  \bibinfo{pages}{024313}, \doiprefix\url{10.1103/PhysRevC.82.024313}
  (\bibinfo{year}{2010}).

\bibitem{UNEDF1}
\bibinfo{author}{Kortelainen, M.} \emph{et~al.}
\newblock \bibinfo{journal}{\bibinfo{title}{Nuclear energy density
  optimization: Large deformations}}.
\newblock {\emph{\JournalTitle{Phys. Rev. C}}} \textbf{\bibinfo{volume}{85}},
  \bibinfo{pages}{024304}, \doiprefix\url{10.1103/PhysRevC.85.024304}
  (\bibinfo{year}{2012}).

\bibitem{UNEDF2}
\bibinfo{author}{Kortelainen, M.} \emph{et~al.}
\newblock \bibinfo{journal}{\bibinfo{title}{Nuclear energy density
  optimization: Shell structure}}.
\newblock {\emph{\JournalTitle{Phys. Rev. C}}} \textbf{\bibinfo{volume}{89}},
  \bibinfo{pages}{054314}, \doiprefix\url{10.1103/PhysRevC.89.054314}
  (\bibinfo{year}{2014}).

\bibitem{massexplorer}
\bibinfo{author}{{Mass Explorer}} (\bibinfo{year}{2020}).
\newblock \bibinfo{note}{\url{http://massexplorer.frib.msu.edu}}.

\bibitem{Moller2012}
\bibinfo{author}{M{\"o}ller, P.}, \bibinfo{author}{Sierk, A.},
  \bibinfo{author}{Ichikawa, T.} \& \bibinfo{author}{Sagawa, H.}
\newblock \bibinfo{journal}{\bibinfo{title}{Nuclear ground-state masses and
  deformations: {FRDM}(2012)}}.
\newblock {\emph{\JournalTitle{At. Data Nucl. Data Tables}}}
  \textbf{\bibinfo{volume}{109-110}}, \bibinfo{pages}{1 -- 204},
  \doiprefix\url{10.1016/j.adt.2015.10.002} (\bibinfo{year}{2016}).

\bibitem{Goriely2013}
\bibinfo{author}{Goriely, S.}, \bibinfo{author}{Chamel, N.} \&
  \bibinfo{author}{Pearson, J.~M.}
\newblock \bibinfo{journal}{\bibinfo{title}{Further explorations of
  {Skyrme-Hartree-Fock-Bogoliubov} mass formulas. {XIII.} the 2012 atomic mass
  evaluation and the symmetry coefficient}}.
\newblock {\emph{\JournalTitle{Phys. Rev. C}}} \textbf{\bibinfo{volume}{88}},
  \bibinfo{pages}{024308}, \doiprefix\url{10.1103/PhysRevC.88.024308}
  (\bibinfo{year}{2013}).

\bibitem{Gneiting2007}
\bibinfo{author}{Gneiting, T.} \& \bibinfo{author}{Raftery, A.~E.}
\newblock \bibinfo{journal}{\bibinfo{title}{Strictly proper scoring rules,
  prediction, and estimation}}.
\newblock {\emph{\JournalTitle{J. Amer. Statist. Assoc.}}}
  \textbf{\bibinfo{volume}{102}}, \bibinfo{pages}{359--378},
  \doiprefix\url{10.1198/016214506000001437} (\bibinfo{year}{2007}).

\bibitem{Raftery2007}
\bibinfo{author}{Gneiting, T.}, \bibinfo{author}{Balabdaoui, F.} \&
  \bibinfo{author}{Raftery, A.~E.}
\newblock \bibinfo{journal}{\bibinfo{title}{Probabilistic forecasts,
  calibration and sharpness}}.
\newblock {\emph{\JournalTitle{J. Roy. Stat. Soc. Ser. B Stat. Methodol.}}}
  \textbf{\bibinfo{volume}{69}}, \bibinfo{pages}{243--268},
  \doiprefix\url{https://doi.org/10.1111/j.1467-9868.2007.00587.x}
  (\bibinfo{year}{2007}).

\bibitem{NUTS}
\bibinfo{author}{Homan, M.~D.} \& \bibinfo{author}{Gelman, A.}
\newblock \bibinfo{journal}{\bibinfo{title}{The {{No-U-Turn Sampler}}:
  Adaptively setting path lengths in {{Hamiltonian Monte Carlo}}}}.
\newblock {\emph{\JournalTitle{J. Mach. Learn. Res.}}}
  \textbf{\bibinfo{volume}{15}}, \bibinfo{pages}{1351--1381}
  (\bibinfo{year}{2014}).

\bibitem{Clyde1996}
\bibinfo{author}{Clyde, M.}, \bibinfo{author}{Desimone, H.} \&
  \bibinfo{author}{Parmigiani, G.}
\newblock \bibinfo{journal}{\bibinfo{title}{Prediction via orthogonalized model
  mixing}}.
\newblock {\emph{\JournalTitle{J. Am. Stat. Assoc.}}}
  \textbf{\bibinfo{volume}{91}}, \bibinfo{pages}{1197--1208},
  \doiprefix\url{10.1080/01621459.1996.10476989} (\bibinfo{year}{1996}).

\bibitem{BDA}
\bibinfo{author}{Gelman, A.} \emph{et~al.}
\newblock \emph{\bibinfo{title}{Bayesian Data Analysis}}
  (\bibinfo{publisher}{CRC Pres}, \bibinfo{year}{2013}),
  \bibinfo{edition}{third} edn.

\bibitem{HFB24dataset}
\bibinfo{author}{Goriely, S.}, \bibinfo{author}{Chamel, N.} \&
  \bibinfo{author}{Pearson, J.~M.}
\newblock \bibinfo{title}{{HFB-24} mass formula} (\bibinfo{year}{2020}).
\newblock
  \bibinfo{note}{\url{http://www.astro.ulb.ac.be/bruslib/nucdata/hfb24-dat}}.

\bibitem{PyMC3}
\bibinfo{author}{Salvatier, J.}, \bibinfo{author}{Wiecki, T.~V.} \&
  \bibinfo{author}{Fonnesbeck, C.}
\newblock \bibinfo{journal}{\bibinfo{title}{Probabilistic programming in python
  using {PyMC3}}}.
\newblock {\emph{\JournalTitle{PeerJ Comp. Sci.}}}
  \textbf{\bibinfo{volume}{2:e55}}, \bibinfo{pages}{1351--1381},
  \doiprefix\url{10.7717/peerj-cs.55} (\bibinfo{year}{2016}).

\end{thebibliography}

\clearpage
\newpage

\section*{Methods}
\setcounter{figure}{0}
\setcounter{table}{0}
\renewcommand{\figurename}{METHODS Figure}
\renewcommand{\tablename}{METHODS Table}
\renewcommand{\thefigure}{\arabic{figure}}
\renewcommand{\thetable}{\Roman{table}}
\renewcommand{\theHfigure}{METHODS Figure \thefigure}
\renewcommand{\theHtable}{METHODS Table \thetable}

\subsection*{BMA highlights}
Let us consider the task of predicting observations from a physical process at new locations $x^*$ using the observations $\yb = (y(x_1), \dots, y(x_{n}))$. The BMA posterior predictive distribution is 
\begin{equation*}\label{eqn:BMA}
	p(y(x^*)|\yb) 
	= \sum_{k=1}^p p(y(x^*)|\yb,\mathcal{M}_k) p(\mathcal{M}_k|\yb).
\end{equation*}
This is simply a linear combination of individual models' posterior predictive distributions. 
The global model weights are taken as 
the posterior probabilities $p(\mathcal{M}_k|\yb)$ 
that the model $\mathcal{M}_k$ is the true model as given by the Bayes' theorem:
\begin{equation*} \label{eqn:posteriorsMmodel:BMA}
p(\mathcal{M}_k|\yb)
= \frac{p(\yb|\mathcal{M}_k)\pi(\mathcal{M}_k)}{\sum_{\ell=1}^p p(\yb|\mathcal{M}_\ell) \pi(\mathcal{M}_\ell)},
\end{equation*}
where
\begin{equation*} \label{eqn:evidencel}
p(\yb|\mathcal{M}_k) =\int p(\yb|\sigma_k,\deltab_{f,k},\mathcal{M}_k)\pi( \sigma_k, \deltab_{f,k}|\mathcal{M}_k) d \sigma_k d \deltab_{f,k}
\end{equation*}
is the evidence (integral) of model $\mathcal{M}_k$ and $\pi( \sigma_k, \deltab_{f,k}|\mathcal{M}_k)$ is the prior density of model's parameters (noise scale $\sigma_k$ and systematic discrepancy $\deltab_{f,k}$), 
$p(\yb|\sigma_k, \deltab_{f,k}, \mathcal{M}_k)$ is the data likelihood, and $\pi(\mathcal{M}_k)$ is the prior probability that $\mathcal{M}_k$ is the true model 
-- assuming that one of the models is true.

There is only a handful of statistical distributions under which the evidence integral can be expressed in a closed form. 
One such scenarios is linear regression models with conjugate priors; the statistical model $\mathcal{M}_k$ with a constant discrepancy term $\delta_{f,k}$ belongs to this case. 
For each model, let us consider the prior 
\begin{equation*}
\pi (\delta_{f,k},\lambda_k| \mathcal{M}_k) = \pi (\delta_{f,k}|\lambda_k, \mathcal{M}_k) \pi(\lambda_k| \mathcal{M}_k),
\end{equation*}
where $\delta_{f,k}$, 
conditionally on 
a theoretical model choice $\mathcal{M}_k$
and a precision parameter 
(the inverse of the variance) 
$\lambda_k$, 
follows a normal distribution with mean $\mu$ and variance $1 / \lambda_k$. 
Let us further assign to the precision $\lambda_k$
a gamma prior with shape parameter $a$ and rate parameter $b$.
Then, the evidence integral has the closed form solution:
\begin{equation*}
\label{eq:evidence-closed-form}
p(\yb| \mathcal{M}_k) = \frac{\Gamma(a_n) b^a} {\Gamma(a)b_n^{a_n}\kappa_n^{\frac{1}{2}}} (2 \pi)^{-\frac{n}{2}},
\end{equation*}
where
$a_n = a + \frac{n}{2}$,
$b_n =b + \frac{1}{2} \sum_{i = 1}^{n} (d_i - \bar{d})^2 + \frac{n (\bar{d} - \mu)^2}{2(1 + n)}$,
$\kappa_n = 1 + n$,
while denoting $d_i := y_i - y_k(x_i)$ and $\bar{d} := (\sum_i d_i) / n$. This solution can be obtained by simple but tedious algebraic manipulations, see Ref.~\cite{BDA} for details. As stated in the main manuscript, we use for the parameter $\sigma$ a gamma prior with scale and rate parameters 5 and 10. 
In order to match the mean and standard deviation 
of $1 / \sigma^2$ when $\sigma$ is distributed according to the common Gamma prior with shape and scale parameters 5 and 10,
the results for the closed form BMA were obtained under a gamma prior for the precision (inverse variance) parameter with shape and scale parameters 0.252 and 0.030.

When evidences cannot be obtained explicitly, 
a MC estimate can be computed as
\begin{equation}
\widehat{p_{MC}(\yb|\mathcal{M}_k)} = \frac{1}{n_{MC}} \sum_{i = 1}^{n_{MC}} p(\yb|\deltab^{(i)}_{f,k},\sigma^{(i)}_k,\mathcal{M}_k),\nonumber
\end{equation}
where $\{(\deltab^{(i)}_{f,k},\sigma^{(i)}_k): i = 1,\dots, n_{MC}\}$ 
are samples from the prior distribution of model parameters $\pi (\deltab_{f,k},\sigma_k| \mathcal{M}_k)$.

Alternatively, 
when the discrepancy term is considered constant, 
the evidence integral can be approximated by
a closed form expression. 
A technique frequently used is Laplace's 
method of integral quadrature
\cite{Hoeting1999}:
\begin{equation}
    \widehat{p_L(\yb|\mathcal{M}_k)} =  2\pi | \widetilde{\Sigma}_k|^{\frac{1}{2}} p(\yb|\tilde{\delta}_{f,k},\tilde{\sigma}_k,\mathcal{M}_k) \pi(\tilde{\delta}_{f,k},\tilde{\sigma}_k|\mathcal{M}_k),\nonumber
\end{equation}
where $\tilde{\sigma}_k$ and $\tilde{\delta}_{f,k}$ represent the posterior modes and $\widetilde{\Sigma}_k = (-\Db^2 l(\tilde{\delta}_{f,k}, \widetilde{\sigma}_k))^{-1}$ is the inverse of the Hessian matrix of second derivatives of $l(\delta_{f,k},\sigma_k) = \log p(\yb|\delta_{f,k}, \sigma_k,\mathcal{M}_k) + \log \pi(\delta_{f,k}, \sigma_k|\mathcal{M}_k)$. 
For $\sigma_k \sim \text{Gamma}(a, b)$ and $\delta_{f,k} \sim N(\mu, s^2)$, we have
\begin{align*}
\frac{\partial^2 l(\delta_{f,k},\sigma_k)}{\partial \sigma^2_k} &= \frac{n-a+1}{\sigma_k^2} - \frac{3\sum_i (y(x_i) - y_k(x_i) - \delta_{f,k})^2}{\sigma_k^4}, \\
  \frac{\partial^2 l(\delta_{f,k},\sigma_k)}{\partial \sigma_k \partial \delta_{f,k}} &= - \frac{2 \sum_i (y(x_i) - y_k(x_i) - \delta_{f,k})}{\sigma_k^3} - \frac{\delta_{f,k} - \mu}{s^2},\\
  \frac{\partial^2 l(\delta_{f,k},\sigma_k)}{ \partial \delta_{f,k}^2} &= - \frac{n}{\sigma_k^2} - \frac{1}{s^2}.
\end{align*}{}
The computation of evidence integral is simplified in the scenarios without the constant discrepancy term $\delta_{f,k}$ as the statistical model contains only a single parameter $\sigma_k$. We leave the details of this simple exercise in probability to the reader.

\subsection*{Application: summary of modeling choices}
As a matter of clarity and to guarantee reproducibility of the results presented in section \textit{Application: nuclear mass exploration}, Table~\ref{tab:Models} lists parameter choices and priors for each of the modeling variants discussed. Note that when we consider theoretical model  without statistical correction, $\delta_{f,k}(x_i) := 0$.

\begin{table}[ht!]
\caption{
Summary of statistical models, their parameters, and priors used in section \textit{Application: nuclear mass exploration.} 
    \label{tab:Models}}
    \centering
    \setlength{\tabcolsep}{3pt}
    \fbox{
\begin{tabular}{l|l|l}
 & Statistical Model & Prior distributions \\
 \hline
BMA & $\mathcal{M}_k: y(x_i) = f_k(x_i)+\delta_{f,k}(x_i) + \sigma \epsilon_i$ & \begin{tabular}[c]{@{}l@{}}$1/ \sigma \sim \text{Gamma}(0.252, 0.030)$, \\ $ \pi(\mathcal{M}_k) = 1/9$\end{tabular} \\
\hline
GBMM+L & $y(x_i) = \sum_{k=1}^p \omega_k \left(f_k(x_i)+\delta_{f,k}(x_i)\right) + \sigma \epsilon_i$ & \begin{tabular}[c]{@{}l@{}}$\sigma \sim \text{Gamma}(5, 10)$,\\ $\omegab \overset{\mathrm{iid}}{\sim} \text{Uniform}(0,1)$ \end{tabular} \\
\hline
GBMM+D & $y(x_i) = \sum_{k=1}^p \omega_k \left(f_k(x_i)+\delta_{f,k}(x_i)\right) + \sigma \epsilon_i$ & \begin{tabular}[c]{@{}l@{}}$\sigma \sim \text{Gamma}(5, 10)$,\\ $\omegab|\alphab \sim \text{Dirichlet}(\alpha)$, \\ $\alphab \overset{\mathrm{iid}}{\sim} \text{Half-Normal}(2^2)$\end{tabular}  \\
\hline
LBMM+GLM & $y(x_i) = \sum_{k=1}^p \omega_k(x_i) \left(f_k(x_i)+\delta_{f,k}(x_i)\right) + \sigma \epsilon_i$ & \begin{tabular}[c]{@{}l@{}}$\sigma \sim \text{Gamma}(5, 10)$,\\ $\omegab(x_i) |\alphab(x_i) \sim \text{Dirichlet}(\alphab(x_i))$, \\ $\log(\alpha_k(x_i)) = \betab^T_k x_i $, \\ $\betab_k \overset{\mathrm{iid}}{\sim} N(0,1)$ \end{tabular} \\
\hline
LBMM+GPM & $y(x_i) = \sum_{k=1}^p \omega_k(x_i) \left(f_k(x_i)+\delta_{f,k}(x_i)\right) + \sigma \epsilon_i$ & \begin{tabular}[c]{@{}l@{}}$\sigma \sim \text{Gamma}(5, 10)$,\\ $\omegab(x_i) |\alphab(x_i) \sim \text{Dirichlet}(\alphab(x_i))$, \\ $\log(\alpha_k(x_i)) \sim \text{GP}(\gamma_k^\infty, c_k(x_i, x_i'))$, where \\ $c_k(x_i, x_i') = \eta_k 
e^{-\frac{(Z_i - Z_i')^2}{2\rho_{Z}^2}
-\frac{(N_i - N_i')^2}{2\rho_{N}^2}}$ and\\
$\gamma_k^\infty \sim  N(0,1)$,\\
$\eta_k \sim \text{Gamma}(10,2)$,\\
$\rho_N \sim \text{Gamma}(5,2)$,\\
$\rho_Z \sim \text{Gamma}(5,2)$
\end{tabular}
\end{tabular}}
\end{table}

\subsection*{MCMC computations} 
\begin{figure*}[htb!]
	\begin{centering}
		\includegraphics[width=0.95\textwidth]{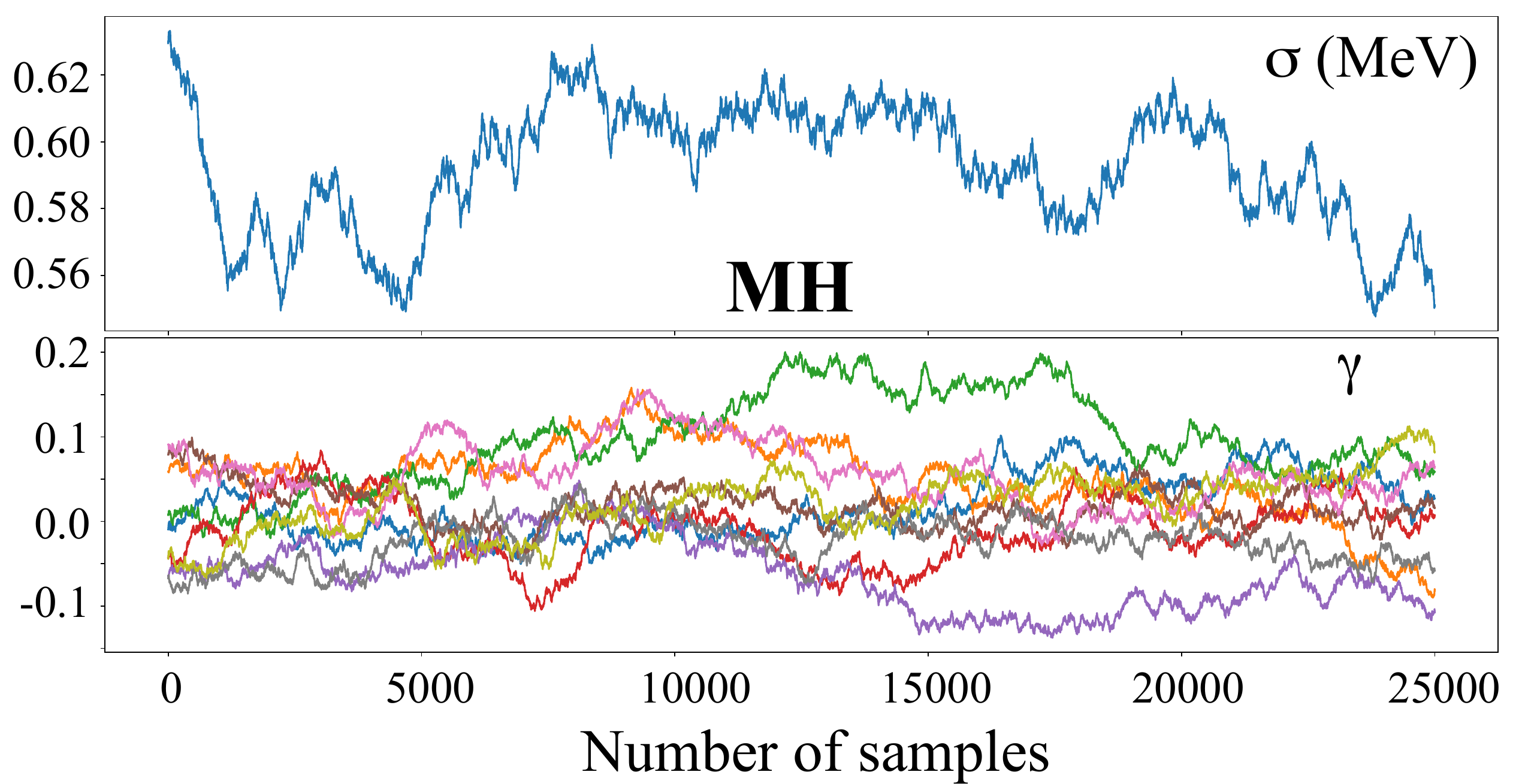}
		\caption{Traceplots of the scale parameter $\sigma$ and GP mean parameters $\gamma_k^\infty$ obtained via the Metropolis-Hastings algorithm in the LBMM+GPD variant.\label{fig:traceplot_MH}}
	\end{centering}
\end{figure*}

\begin{figure*}[htb!]
	\begin{centering}
		\includegraphics[width=0.95\textwidth]{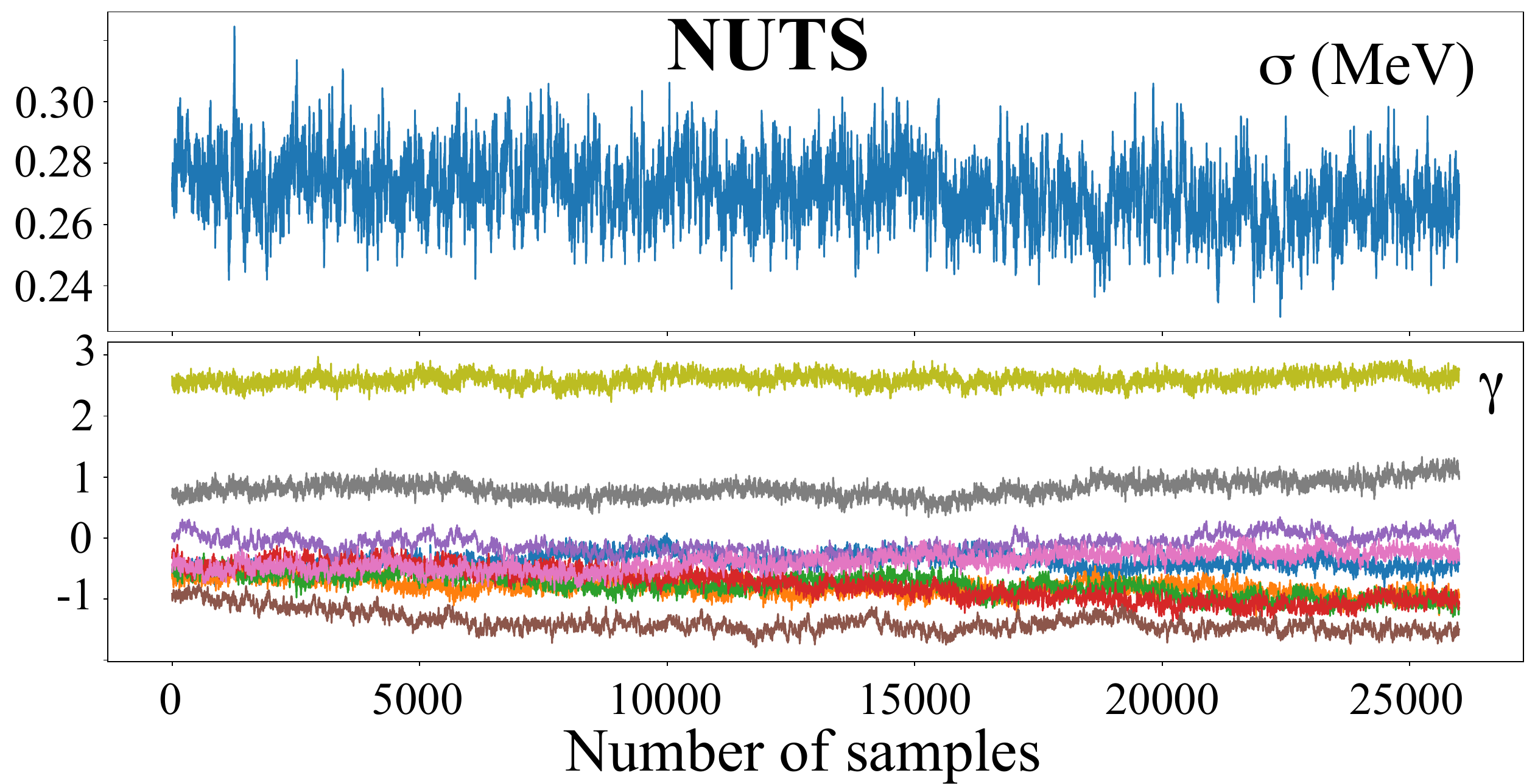}
		\caption{Similar as in METHODS Fig.~\ref{fig:traceplot_MH} but for the No-U-Turn sampler.\label{fig:traceplot_NUTS}}
	\end{centering}
\end{figure*}

The MCMC approximate posterior distributions for all the modeling variants discussed in this work were obtained using the Hamiltonian Monte Carlo based No-U-Turn sampler (NUTS) \cite{NUTS}. In general, we obtained at least $50 \times 10^3$ samples from the posterior distributions after which we discarded half as a burn-in. While more conventional samplers such as Metropolis-Hastings (MH) algorithm \cite{BDA} would be sufficient for both BMA and global mixtures, using NUTS is essential to achieve satisfactory convergence when it comes to LBMM. To illustrate this, we provide selected MCMC traceplots for LBMM+GPD variant with MH and NUTS approximations in METHODS Figs. \ref{fig:traceplot_MH} and \ref{fig:traceplot_NUTS}, respectively. The NUTS whose performance tends to be superior to MH in scenarios with moderately large parameter spaces clearly shows the convergence of Markov Chain while the MH displays poor mixing.

\section*{Data availability}
 The experimental data used in this study comes from the publicly available measurements collected in 
 AME2003 \cite{AME03b} and AME2020 \cite{AME2020a}. 
 The results of the nuclear DFT mass models are publicly available 
 from the theoretical database MassExplorer \citemethods{massexplorer}. 
The  FRDM-2012 results were taken from the supplementary data of Ref. \citemethods{Moller2012}. The  HFB-24 mass predictions were taken from the Brusslib website \cite{HFB24dataset}.
\section*{Code availability}
MCMC sampling from the Bayesian posterior distributions was performed with the Python package PyMC3 (version 3.3)\cite{PyMC3}. The Bayesian calculations of the theoretical binding energies that support the findings of this study 
will be made publicly available in \url{https://github.com/kejzlarv} upon publication of this article.

\section*{Supplementary Information} 
Supplementary information is linked to the online version of the paper at https://www.nature.com/srep/.

\section*{Acknowledgements}
Useful comments from R.J. Furnstahl are gratefully acknowledged.
This material is based upon work supported by the U.S.
Department of Energy, Office of Science, Office of Nuclear
Physics under Awards Nos. DE-SC0023688  and DOE-DE-SC0013365, and
by the National Science Foundation under award number 2004601 (CSSI program, BAND collaboration).

\section*{Author contributions}
V.K. and L.N. performed the Bayesian analysis. 
All authors discussed the results and prepared the manuscript.

\section*{Author information} 
Reprints and permissions information is available at www.nature.com/reprints.
The authors declare no competing financial interests. Readers are welcome to
comment on the online version of the paper. Correspondence and requests for
materials should be addressed to W.N. (witek@frib.msu.edu)

\section*{Competing interests statement}
The authors declare that they have no competing financial interests.

\end{document}